\documentclass[12pt,a4paper,twoside,english]{iopart}
\usepackage[T1]{fontenc}
\usepackage[latin9]{inputenc}
\usepackage{color}
\usepackage{units}
\usepackage{textcomp}
\usepackage{amstext}
\usepackage{amssymb}
\usepackage{graphicx}
\usepackage[numbers]{natbib}

\makeatletter

\pdfpageheight\paperheight
\pdfpagewidth\paperwidth

\newcommand{\lyxmathsym}[1]{\ifmmode\begingroup\def\b@ld{bold}
  \text{\ifx\math@version\b@ld\bfseries\fi#1}\endgroup\else#1\fi}

\usepackage{iopams}
\usepackage{setstack}

\newcommand{\eqref}[1]{(\ref{#1})}

\makeatother

\usepackage{babel}
\begin{document}

\title{Superconducting Atomic Contacts inductively coupled to a microwave
resonator }

\author{\noindent C. Janvier$^{1}$, L. Tosi$^{2}$, Ç. Ö. Girit$^{1}$,
M. F. Goffman$^{1}$, H. Pothier$^{1}$, C.~Urbina$^{1}$.}

\address{\noindent $^{1}$Quantronics Group, Service de Physique de l'État
Condensé (CNRS, URA\ 2464), IRAMIS, CEA-Saclay, 91191 Gif-sur-Yvette,
France}

\address{\noindent $^{2}$Centro Atómico Bariloche and Instituto Balseiro,
Comisión Nacional de Energía Atómica (CNEA), 8400 Bariloche, Argentina}

\ead{\noindent cristian.urbina@cea.fr}
\begin{abstract}
\noindent We describe and characterize a microwave setup to probe
the Andreev levels of a superconducting atomic contact. The contact
is part of a superconducting loop inductively coupled to a superconducting
coplanar resonator. By monitoring the resonator reflection coefficient
close to its resonance frequency as a function of both flux through
the loop and frequency of a second tone we perform spectroscopy of
the transition between two Andreev levels of highly transmitting channels
of the contact. The results indicate how to perform coherent manipulation
of these states.
\end{abstract}


\maketitle

\section{Introduction}

Atomic-size contacts between metallic electrodes are routinely obtained
using either scanning tunneling microscopes or break-junctions \citep{Agrait}.
From the electrical transport point of view, atomic contacts are simple
systems. First, as for any good metal, electron-electron interactions
are strongly screened. Second, because their transverse dimensions
are of the order of the Fermi wavelength (typically $0.2\,\mathrm{nm}$)
they accommodate only a small number of conduction channels. Moreover,
as the transmission probability $\tau_{i}$ for electrons through
each conduction channel can be adjusted and measured in-situ \citep{Scheer},
atomic contacts provide a test-bed to explore mesoscopic electronic
transport \citep{Nazarov Blanter,Cron shot noise,Cron DCB}. In particular,
when the metal is a superconductor atomic-contacts constitute elementary
Josephson weak-links that allow probing the foundations of the Josephson
effect \citep{Bretheau CRAS}.

Josephson supercurrents~\citep{Josephson} will flow through any
barrier weakly coupling two superconductors, including a tunnel junction,
a constriction, a molecule, or a normal metal \citep{Golubov}. Microscopically,
weak links differ in their local quasiparticle excitation spectrum.
For a non-interacting system, this spectrum is determined by the length
of the weak link, compared to the superconducting coherence length,
and its configuration of conduction channels as characterized by the
set of transmission probabilities $\left\{ \tau_{i}\right\} $. The
excitation spectrum of a short single-channel weak link of arbitrary
transmission $\tau$ contains, besides the continuum of states at
energies larger than the superconducting gap $\Delta,$ a sub-gap
spin-degenerate level, known as the Andreev doublet (Figure~\ref{Fig. Andreev}).
Its energy $E_{A}=\Delta\sqrt{1-\tau\sin^{2}\left(\delta/2\right)}$
\citep{Furusaki,Beenakker} is a $2\pi-$periodic function of the
superconducting phase difference $\delta$ across the weak link. It
is precisely this phase dependence that gives rise to the Josephson
supercurrent. In the widespread case of Josephson tunnel junctions,
for which all $\tau_{i}\ll1$, $E_{Ai}\sim\Delta$ and the lowest-lying
excitations conserving electron parity have a threshold energy only
slightly lower than $2\Delta$. By absorbing energy $\gtrsim2\Delta$
a pair can be broken and two quasiparticles created at essentially
the gap energy $\Delta,$ like in a bulk superconductor.

\begin{center}
\begin{figure}[h]
\centering{}\includegraphics[height=130pt]{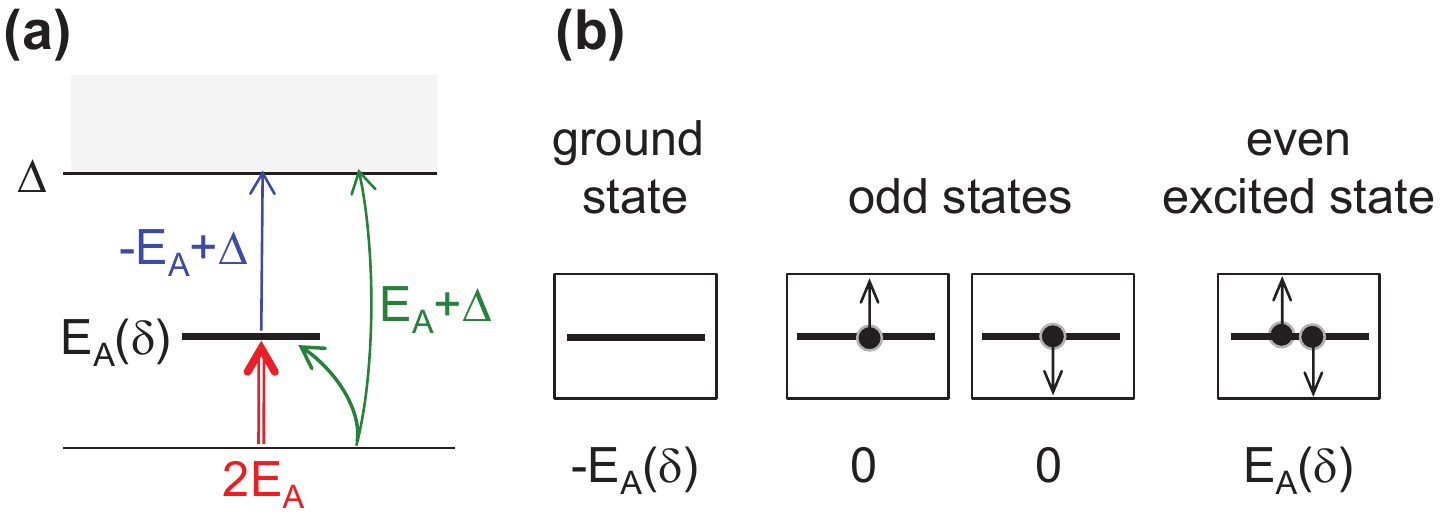}\label{Fig. Andreev}\caption{\textbf{(}\textbf{\textcolor{black}{a)}} The single particle excitation
spectrum for each channel of a short Josephson weak link consists
of a doubly-degenerate Andreev level at energy $E_{A}\left(\delta\right),$
and a continuum of states at energies larger than the superconducting
gap $\Delta.$ Arrows indicate transitions that can be induced by
microwave absorption. The four possible occupations of the Andreev
level are shown in\textbf{ (}\textbf{\textcolor{black}{b)}}: they
correspond to the ground state (Andreev level empty), the two odd
states (a single quasiparticle in the Andreev level), and the excited
pair state (doubly occupied Andreev level), with energies $-E_{A},$
$0$ and $E_{A}$ respectively.}
\end{figure}

\par\end{center}

Two different spectroscopy experiments have recently probed this excitation
spectrum for superconducting atomic contacts containing channels of
high transmisssion. The first experiment \citep{Nature Bretheau,Bretheau Thesis}
spotlighted the lowest energy excitation that conserves electron parity,
the ``Andreev transition'' of energy $2E_{A}$, which leaves two
quasiparticles in the Andreev level (red double arrow in Figure~\ref{Fig. Andreev}a).
The second experiment \citep{PRX Bretheau} revealed a second type
of excitation, with minimum energy $E_{A}+\Delta$. In this case,
a localized Andreev pair is broken into one quasiparticle in the Andreev
level and one in the continuum (green arrows in Figure~\ref{Fig. Andreev}a)
\citep{Bergeret,Bergeret2,Kos}, thus leaving the system in an ``odd''
state. These odd states had been previously detected through the spontaneous
trapping of a single out-of-equilibrium quasiparticle in the Andreev
doublet {[}17{]}. This ensemble of results firmly support the description
of the Josephson effect in terms of Andreev bound states.

If parity is conserved, the ground state and the even excited state
constitute a two-level system \citep{Ivanov} that has been proposed
as the basis for a new class of superconducting qubits \citep{Lantz2002,Desposito2001,Zazunov2003,Zazunov2005}.
What is particularly interesting and novel is that in contrast with
all other superconducting qubits based on Josephson junction circuits
\citep{Wendin2007} an Andreev qubit is a microscopic two-level system
akin to spin qubits in semiconducting quantum dots. Also, if one considers
the odd states, despite the absence of actual barriers the system
can be viewed as a superconducting ``quantum dot'' allowing manipulation
of the spin degree of freedom of a single quasiparticle \citep{Chtchelkatchev,Michelsen,Paduradiu2012}.
The coherence properties of Andreev doublets {[}20, 22{]} are still
to be addressed experimentally. The relaxation time of the excited
state and the dephasing time of quantum superpositions of the two
states have to be measured, understood, and if possible, controlled.

Both relaxation and dephasing mechanisms contribute in principle to
the linewidth of the Andreev transition. In order to achieve coherent
manipulation of these Andreev states one would need much narrower
lines than those observed in the aforementioned experiments, where
they were typically larger than 500MHz. This was most probably due
to large superconducting phase fluctuations imposed by the dissipative
measurement lines that were necessary to measure the current-voltage
characteristics of the contacts, a key piece of information from which
the $\left\{ \tau_{i}\right\} $ are extracted. Here, to isolate efficiently
the contact from external perturbations, we follow a strategy that
has been implemented successfully for superconducting qubits \citep{Blais2004,Blais2007}.
The idea is to include an atomic contact in a small superconducting
loop to form an rf-SQUID inductively coupled to the electromagnetic
field of a coplanar microwave resonator. The latter should act as
a narrow-band filter to allow probing the Andreev transition at $2E_{A}$
without excessive broadening. A similar setup was analyzed theoretically
in \citep{Romero2012}, although here we have in addition avoided
any galvanic connection of the SQUID loop with the rest of the circuit
in order to minimize the probability of trapping out-of-equilibrium
quasiparticles in the contact {[}17{]}. By varying the flux threading
the SQUID loop the Andreev transition frequency can be brought into
resonance with the resonator mode. This will result in hybridization
of the Andreev levels and the cavity mode (see Figure~\ref{Fig: cavity}).
The goal of the experiment presented here is to perform spectroscopy
of the Andreev levels of the contact as a first step towards coherent
manipulation of the two-level system. 

\begin{center}
\begin{figure}[h]
\begin{centering}
\includegraphics[clip,width=6cm]{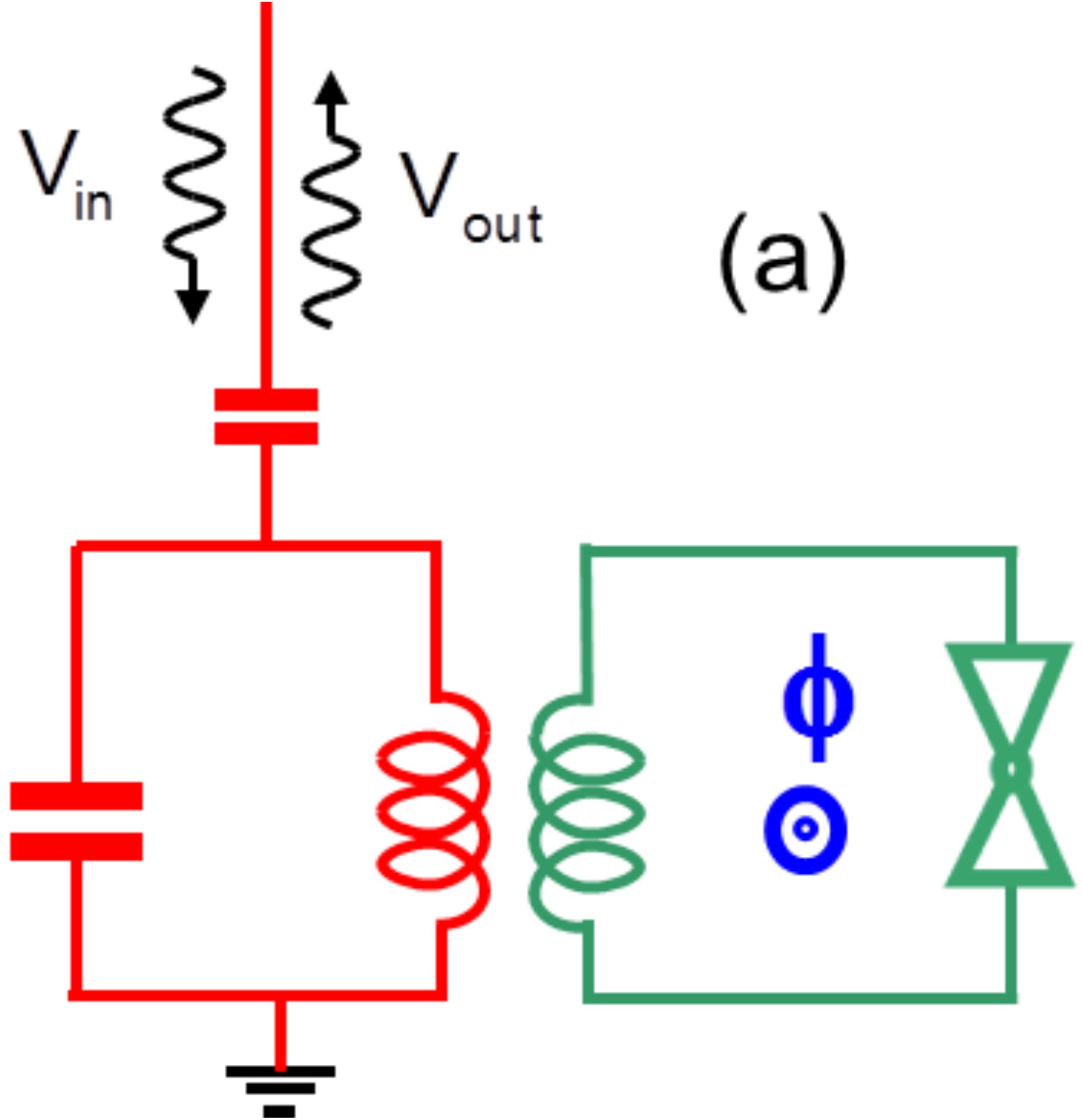}\includegraphics[width=8.4cm]{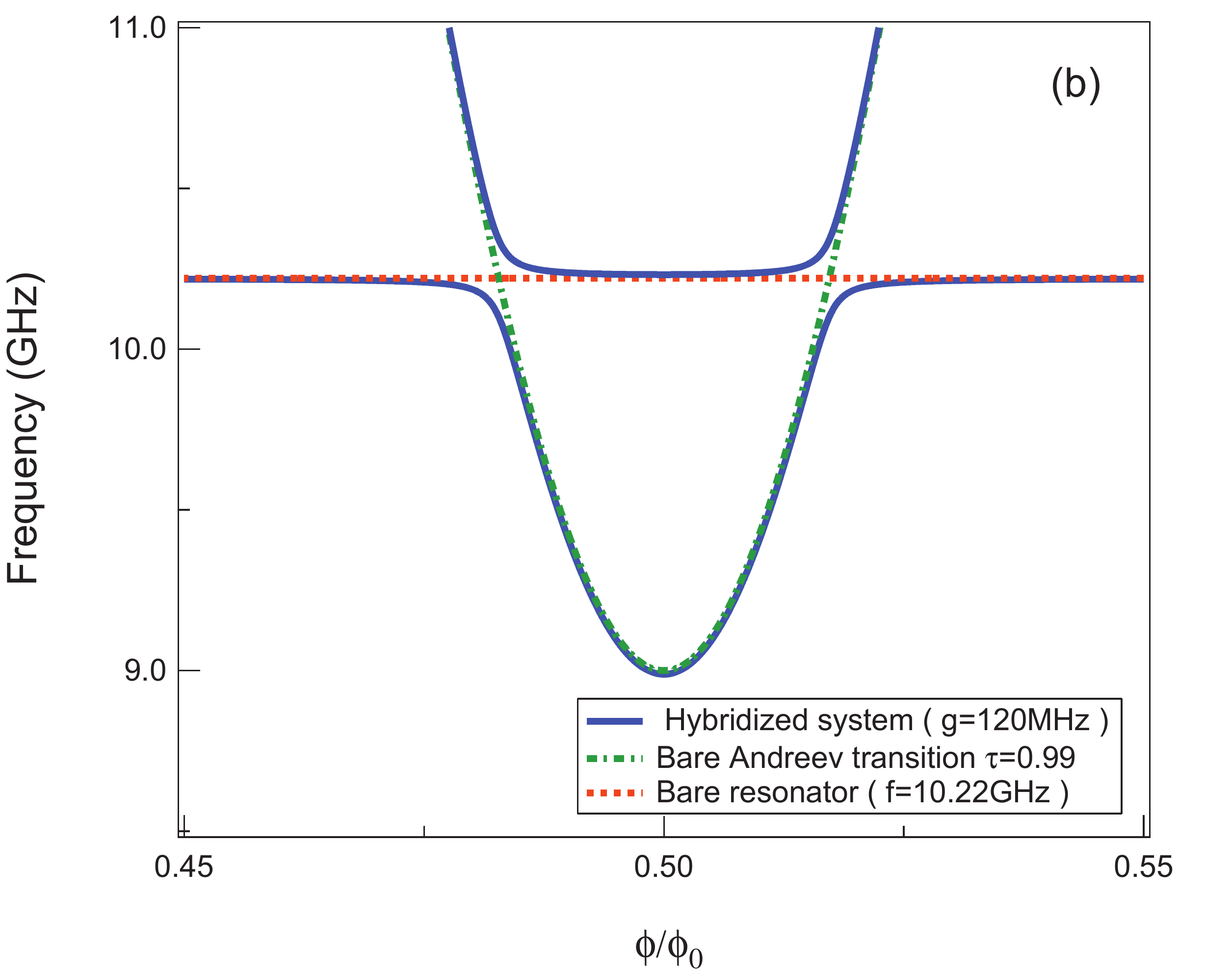}
\par\end{centering}

\caption{\textbf{(a)} An atomic rf-SQUID (in green) inductively coupled to
a microwave resonator, represented here by a LC circuit (in red).
The double triangle symbol represents the atomic contact. The spectrum
of the combined system is probed through microwave reflectometry by
weakly coupling the resonator to the external setup through a small
capacitor. \textbf{(b)} example of the expected spectrum (blue full
lines) as a function of the magnetic flux threading the SQUID loop.
The resonance frequency of the bare resonator (red dashed line) is
here $10.1\,\mathrm{GHz}$; the Andreev frequency (green dashed line)
corresponds to a channel with $\tau=0.991;$ and the SQUID-resonator
coupling energy is $h\times\left(120\,\mathrm{MHz}\right)$. The anti-crossing
results from the hybridization of the two quantum systems.}
\label{Fig: cavity}
\end{figure}

\par\end{center}

\section{Experimental Methods}

\subsection{Sample fabrication}

The samples are fabricated on a flexible $500\,\mathrm{\text{\textmu m-}}$thick
Kapton substrate ($\epsilon_{r}\simeq3.2$,\textcolor{black}{{} $\tan\delta\sim1\times10^{-4}$
at $30\,\mathrm{mK}$ }), $50\,\mathrm{mm}$ in diameter. In a first
step, a series of $\lambda/4$ Nb resonators is fabricated. The substrate
is then cut into $7\,\mathrm{mm\times16\,\mathrm{mm}}$ chips which
are individually processed afterward to fabricate the atomic SQUID.

\subsubsection{The microwave resonator}

A $200\,\mathrm{nm}$ thick Nb layer sputtered over the whole substrate
is patterned via optical lithography, and then structured using reactive
ion etching into a series of quarter-wave ($\lambda/4$) resonators
in a coplanar waveguide geometry (see Figure\,\ref{Fig Sample}).
The total length of the $36\,\mathrm{\lyxmathsym{\textmu}m}$ wide
inner conductor is $5\,\mathrm{mm}$. The gap between the inner conductor
and the ground plane is $18\,\mathrm{\lyxmathsym{\textmu}m}$. The
impedance of the coplanar waveguide is $Z_{r}\sim80\,\mathrm{\Omega.}$
The resonator is coupled through an interdigitated capacitor $C\sim5\,\mathrm{fF}$
to a $50\mathrm{\,\Omega}$ line to measure its reflection coefficient
$S_{11}\equiv20\log\left(\nicefrac{V_{out}}{V_{in}}\right).$ \textcolor{black}{For
the 5\,fF coupling capacitor the external losses should dominate
over the internal ones (arising essentially from dielectric losses
in the kapton substrate) leading to a global quality factor on the
order of 1000.}

\begin{center}
\begin{figure}[h]
\begin{centering}
\includegraphics[width=15.5cm]{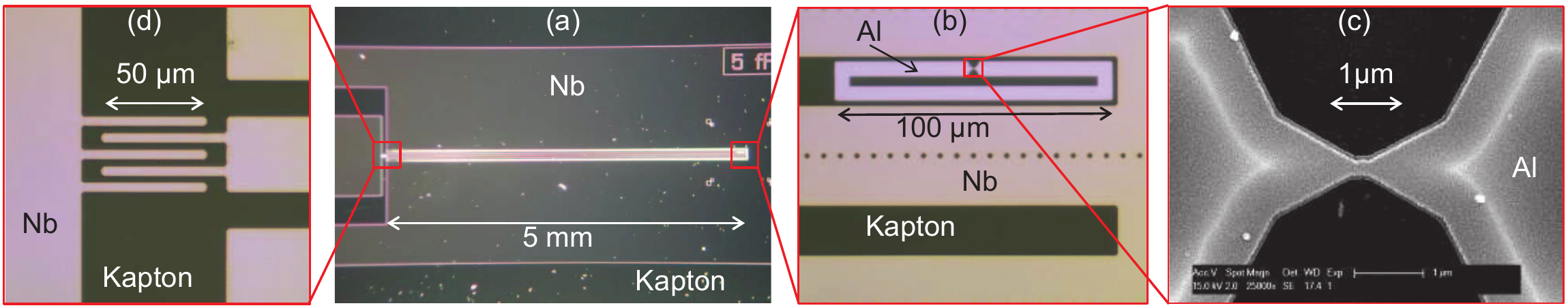}
\par\end{centering}

\caption{\textbf{(a)} Microphotograph of a coplanar quarter-wave Nb resonator.
\textbf{(b)} Zoom on the shorted-end of the microwave Nb resonator
(light gray). The $36\,\mathrm{\text{\textmu m}}$-wide center line
contains some small holes intended as pinning centers for eventual
vortices trapped in the superconducting film. The aluminum loop (white),
with a suspended microbridge in one arm, is placed within the $18\,\mathrm{\lyxmathsym{\textmu}m}$
gap (black). \textbf{(c) }Scanning electron microscope image of suspended
microbridge with a $300\,\mathrm{nm}$-wide constriction. The bright
v-shaped ridges on both sides correspond to the border of the underlying
polyimide layer which was etched to free the bridge over $\sim2\,\mathrm{\text{\textmu m}}$.\textbf{
(d)} interdigitated coupling capacitor of the resonator.}
\label{Fig Sample}
\end{figure}

\par\end{center}

\subsubsection{The atomic rf-SQUID.}

Using electron beam lithography we fabricate a $100\,\mathrm{nm}$-thick
aluminum superconducting loop containing in one arm a micro-bridge,
suspended over approximately $2\,\mathrm{\text{\textmu m}}$ by reactive
ion etching of a sacrificial polyimide layer (Figure\,\ref{Fig Sample}).
The bridge has a $300\,\mathrm{nm}$ constriction at the center. The
width and the inner dimensions of the loop are $5\,\mathrm{\lyxmathsym{\textmu}m}$
and $4\mathrm{\,\lyxmathsym{\textmu}m\times}90\,\mathrm{\lyxmathsym{\textmu}m}$,
respectively, which lead to a geometrical inductance of around $100\,\mathrm{pH}$
, much smaller than the Josephson inductance of a typical atomic contact
(a few nH). A magnetic flux $\phi$ through the loop is then used
to impose a superconducting phase difference $\delta\cong2\pi\nicefrac{\phi}{\phi_{0}}$
across the contact, where $\phi_{0}=\nicefrac{h}{2e}$ is the flux
quantum. This allows adjusting the phase-dependent energy of the Andreev
levels.

\subsection{Setup}

Figure~\ref{BJsetup} shows the break-junction setup. The ensemble
is attached to the mixing chamber of a dilution refrigerator. A precision
screw (not shown), driven by a room temperature dc motor, controls
the vertical displacement of a spindle. A copper slab attached to
the spindle pushes the free end of the sample, which is firmly clamped
on the opposite side against a microwave SMA launcher. The elongation
of the upper face of the substrate as it bends leads eventually to
the bridge rupturing. Afterward, the distance between the two resulting
electrodes varies by a few tens of picometers for every micrometer
of vertical displacement of the pusher.\textbf{ }The temperature of
the ensemble is below $100\,\mathrm{mK}$, and the cryogenic vacuum
ensures that there is no contamination of the freshly exposed electrodes.
The electrodes are gently brought back together, reforming the bridge
and creating an atomic-size contact. Contacts can be made repeatedly
in order to vary the number of channels and/or the transmission probabilities.
An important feature of the microfabricated break junctions \citep{Jan MBJ}
is that a given contact can be maintained for weeks with changes in
transmission below one part in a thousand. 

The sample holder is enclosed in a set of three cylindrical shields
(Al, Cryoperm and Cu, innermost to outermost) attached to the mixing
chamber of a dilution refrigerator. All shields are capped at both
ends. The inside diameter of the Al shield is $76\,\mathrm{mm}$.
The intermediate Cryoperm shield diverts the ambient magnetic field
to reduce flux fluctuations through the SQUID loop as well as to minimize
the number of vortices trapped in the Nb superconducting film. The
inner walls of the Al shield are covered with a $3\,\mathrm{mm}$
thick layer of epoxy loaded with bronze and carbon powder to damp
cavity resonances and adsorb spurious infrared radiation \citep{Barends,Corcoles}.
A small superconducting coil, placed $2\,\mathrm{mm}$ above the sample
inside a copper shield, allows controlling the flux through the loop.

\begin{center}
\begin{figure}[h]
\begin{centering}
\includegraphics[width=8.4cm]{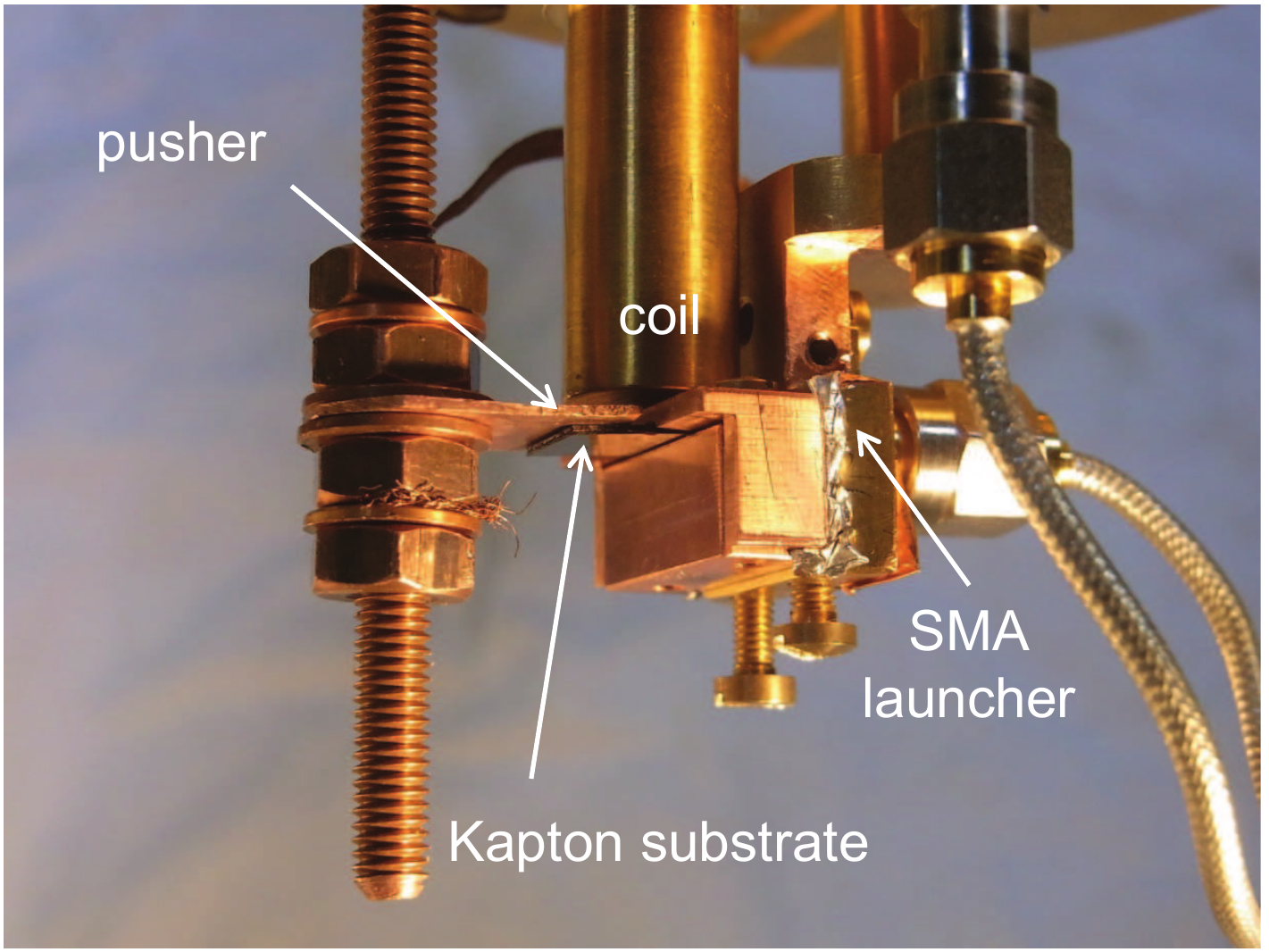}
\par\end{centering}

\caption{Break junction setup: a Kapton substrate is attached to SMA launcher
(right side). The threaded spindle on the left side actuates the pusher
vertically on the free end of the substrate with micrometer precision.
The central cylindrical copper shield hosts a small superconducting
coil that controls the flux through the SQUID loop. The whole system
is enclosed in a set of three shields and anchored to the mixing chamber
of a dilution refrigerator.}
\label{BJsetup}
\end{figure}

\par\end{center}

A single coaxial line enters this set of shields and connects to the
SMA launcher. The overall microwave setup is sketched in Figure\,\ref{Microwave setup}.
There are two 8-12GHz circulators (Pamtech XTE0812KCSD) and one $0.1-18\,\mathrm{GHz}$
directional-coupler (Clear Microwaves C20218) placed at the same temperature
($25\,\mathrm{mK}$) as the sample but outside the shields. A first
microwave tone, injected at one circulator, probes the response of
the resonator at a frequency close to $\nu_{0}$. The reflected signal
from the resonator goes through the two circulators into a cryogenic
amplifier (0.5-11~GHz LNA \#265D from Caltech, gain $28\mathrm{\, dB}$)
placed at 1K. To minimize losses in the return signal a superconducting
coaxial cable (Coax-Co SC-086-50-NbTi-NbTi) is used between the second
circulator and the cryogenic amplifier. The output line from the cryogenic
amplifier to room temperature is a CuNi coax, with a silver cladded
inner conductor (CoaX-Co SC-086/50-SCN-CN). The two circulators prevent
noise from the amplifier reaching the sample. A second tone at frequency
$\nu_{1}$ can be injected through the directional-coupler (-20dB
coupling) to drive the transition between the Andreev levels at the
atomic contact. Each line has a series of attenuators placed at different
stages of the refrigerator to prevent external noise from reaching
the sample. The total attenuation of each of the two input lines (including
losses in the cables) is 90dB.

\begin{center}
\begin{figure}[h]
\centering{}\includegraphics[width=12cm]{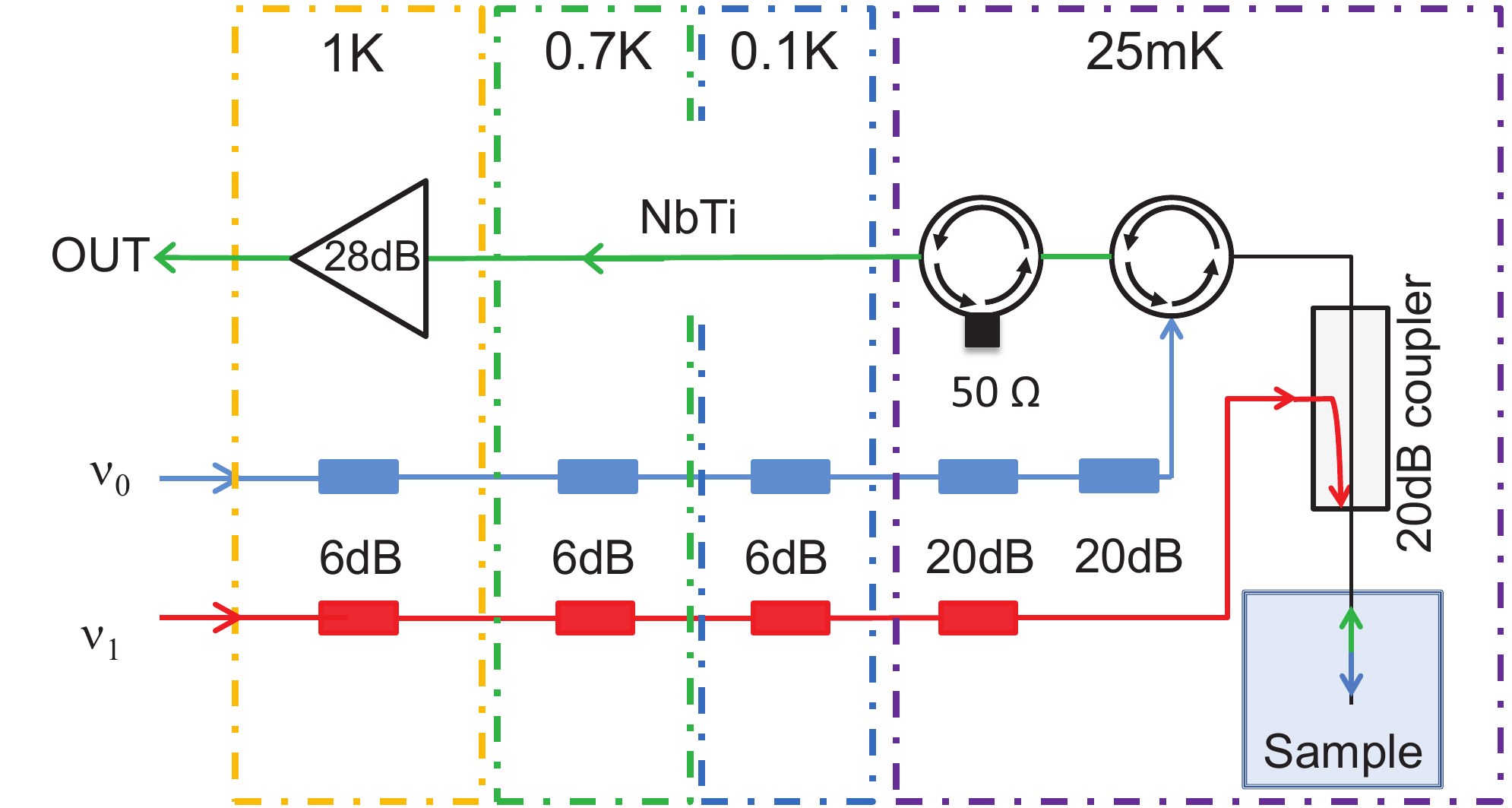}\caption{Microwave setup: A first tone of frequency $\nu_{0}$ is used to probe
the resonator. The two circulators divert the reflected signal towards
a 28dB amplifier placed at 1K. A second tone of frequency $\nu_{1}$
is used to drive the Andreev transition of the atomic contact.}
\label{Microwave setup}
\end{figure}

\par\end{center}

\section{Results}

\subsection{One-tone spectroscopy}

After an additional $78\mathrm{dB}$ room temperature amplification
of the reflected signal, the reflection coefficient is directly measured
using a vector network analyzer. As shown in Figure~\ref{Resonances},
three resonances appear in $S_{11}$ below $10\mathrm{K}$ in the
range $8-10\,\mathrm{GHz}$. Below the superconducting transition
of aluminum, only the two lowest frequency lines (\#1 and \#2) depend
on temperature and bending of the substrate, and are thus associated
with on-chip modes. The coplanar mode resonance is the one at $\nu_{R}\sim10.24\,\mathrm{GHz}$,
with a quality factor $Q\sim300.$ \textcolor{black}{This is three
times lower than expected. The measurements clearly indicate an undercoupled
regime with only 40\textdegree{} phase shift at resonance instead
of the full 360\textdegree{}. We interpret this result as arising
from the coupling of the coplanar resonator mode with a parasitic
mode of the on-chip ground plane which is itself heavily damped by
radiation to the enclosing dissipative cavity }%
\footnote{\textcolor{black}{Because the sample must bent, the two outer electrodes
of the coplanar resonator are actually grounded only at one end. As
a result the ground plane behaves as an antenna. After carrying out
the measurements we understood, through detailed electromagnetic simulations,
that for the actual geometry of the resonator the quarter-wavelength
mode had a similar resonance frequency than the antenna mode of the
ground plane. The two modes hybridize and the resonator mode is also
affected by radiation damping. Hybridization could be avoided by redesigning
the resonator as a meander line (thus making the length of the ground
plane electrode much smaller than the length of the resonator). }%
}\textcolor{black}{.}

\begin{center}
\begin{figure}[h]
\noindent \begin{centering}
\includegraphics[width=12cm]{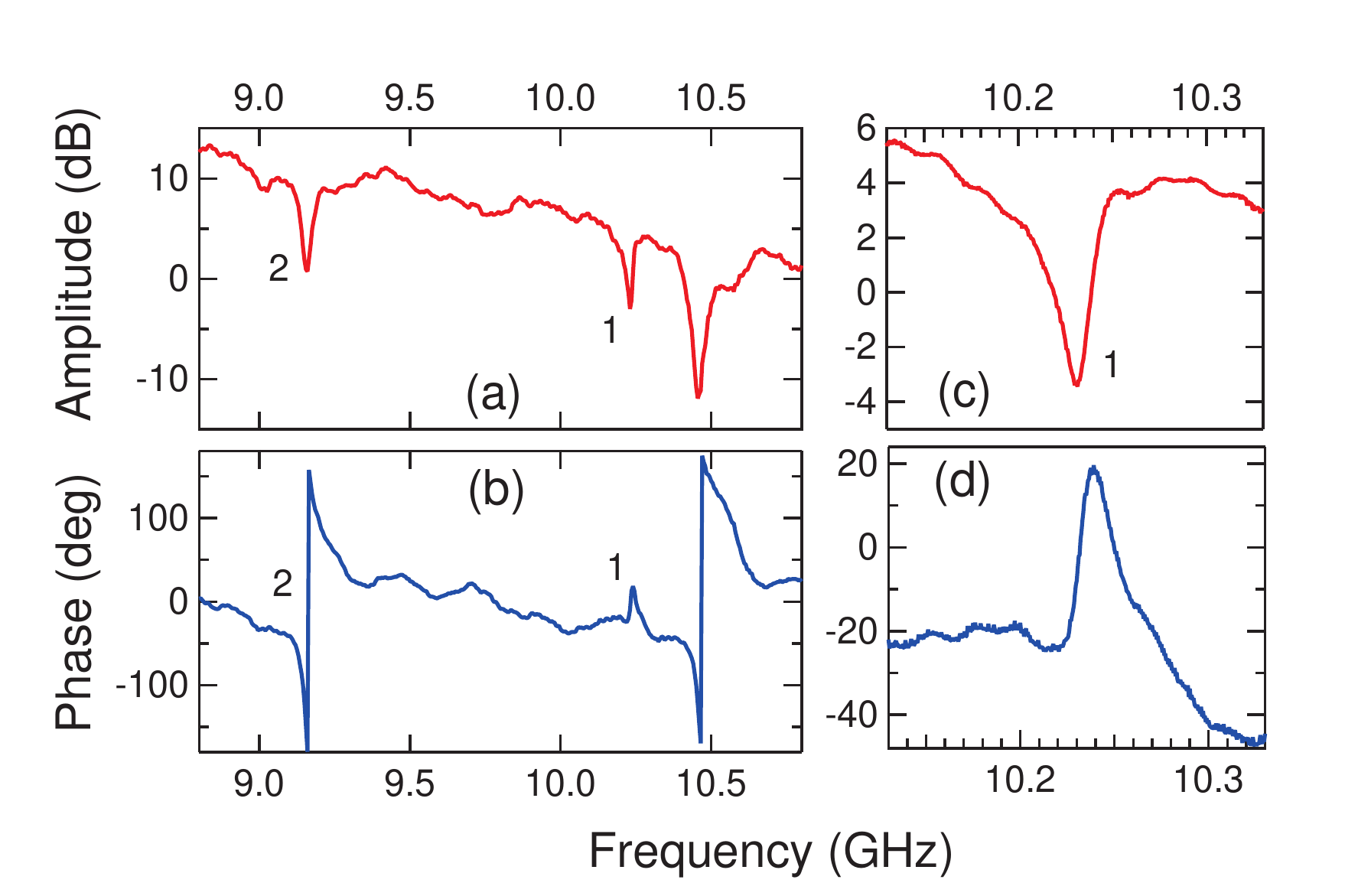}
\par\end{centering}

\centering{}\caption{Amplitude (a) and phase (b) of reflection coefficient of resonator
as a function of frequency, measured at 30~mK. The coplanar mode
resonance is the one at $\nu_{R}\sim10.24\mathrm{GHz}$ (labeled \#1).
There are two parasitic resonances, at 9.16~GHz and 10.45~GHz. Only
resonances \#1 and \#2 change with temperature below 1K and when bending
the substrate and are thus associated to on-chip modes. (c) and (d):
zooms around coplanar mode resonance.}
\label{Resonances}
\end{figure}

\par\end{center}

Despite the low quality factor, it is still possible to probe the
atomic SQUID. Figure~\ref{opening}a shows the evolution of the resonance
frequency of the resonator as the substrate is slowly bent at low
temperature. As the coplanar resonator elongates, its resonance frequency
decreases with the pusher vertical position at a rate of approximately
$50\,\mathrm{kHz}/\mathrm{\text{\textmu m}}$. The sharp frequency
drop observed around a vertical deflection of $400\,\mathrm{\text{\textmu}m}$
signals the last stage of rupture of the break junction. The frequency
shift is in agreement with the change in inductance expected when
opening the SQUID loop. As the vertical displacement of the pusher
is actually not measured in-situ, but deduced from the measured number
of turns of the motor and the pitch of the screw, the backlash of
the mechanical driving setup leads to a hysteresis of around $25\,\mathrm{\text{\textmu}m}$
between opening and closing directions. However, the position at which
the abrupt frequency shift occurs is reproducible within a few microns
for successive openings. In the region of this frequency drop the
contact has atomic dimensions and its Josephson inductance becomes
much larger than the geometrical inductance of the loop. In this limit,
the resonator frequency evolves periodically with the magnetic flux
threading the Al loop, as shown in Figure~\ref{opening}b. If the
substrate is bent further, the loop opens completely and the flux
modulation disappears.

\begin{center}
\begin{figure}[h]
\begin{centering}
\includegraphics[width=12cm]{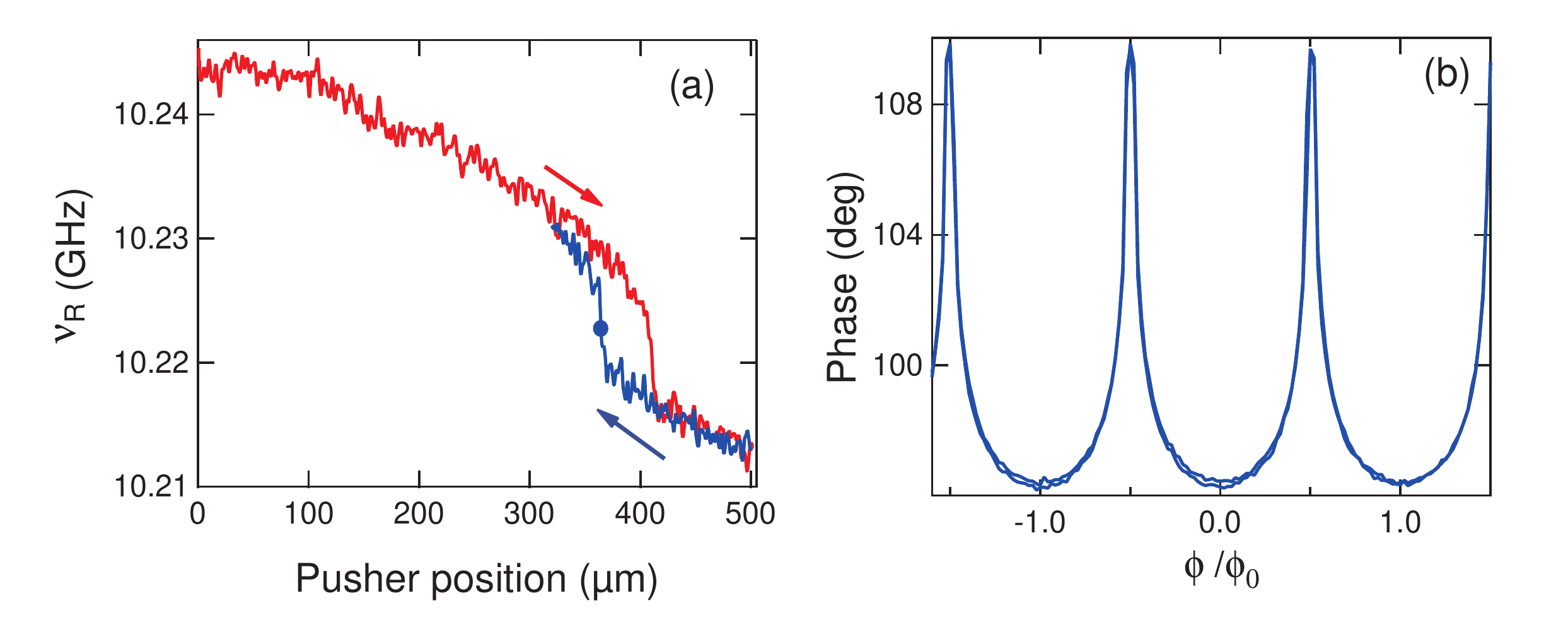}
\par\end{centering}

\caption{(a) Resonance frequency of resonator $\nu_{R}$ as a function of pusher
position. Red (blue) curve correspond to the opening (closing) of
the microbridge. (b) Phase of reflection signal as a function of the
flux threading the SQUID loop (in reduced units) at the pusher position
signaled by the blue dot in right panel. The modulation disappears
when the microbridge is broken.}
\label{opening}
\end{figure}

\par\end{center}

Figure~\ref{Full one-tone spectrum} displays a spectrum of the reflection
coefficient $S_{11}$ as a function of the probe frequency and the
flux threading the loop. An anti-crossing between the resonator and
an Andreev transition in the contact is clearly observed. As shown,
the shape of the spectrum can be described, at least qualitatively,
by considering a single channel of transmission ($\tau\sim0.995$)
coupled ($g\sim120\,\mathrm{MHz})$ to the coplanar resonator harmonic
oscillator \citep{Romero2012}.

\begin{center}
\begin{figure}[h]
\begin{centering}
\includegraphics[width=8.4cm]{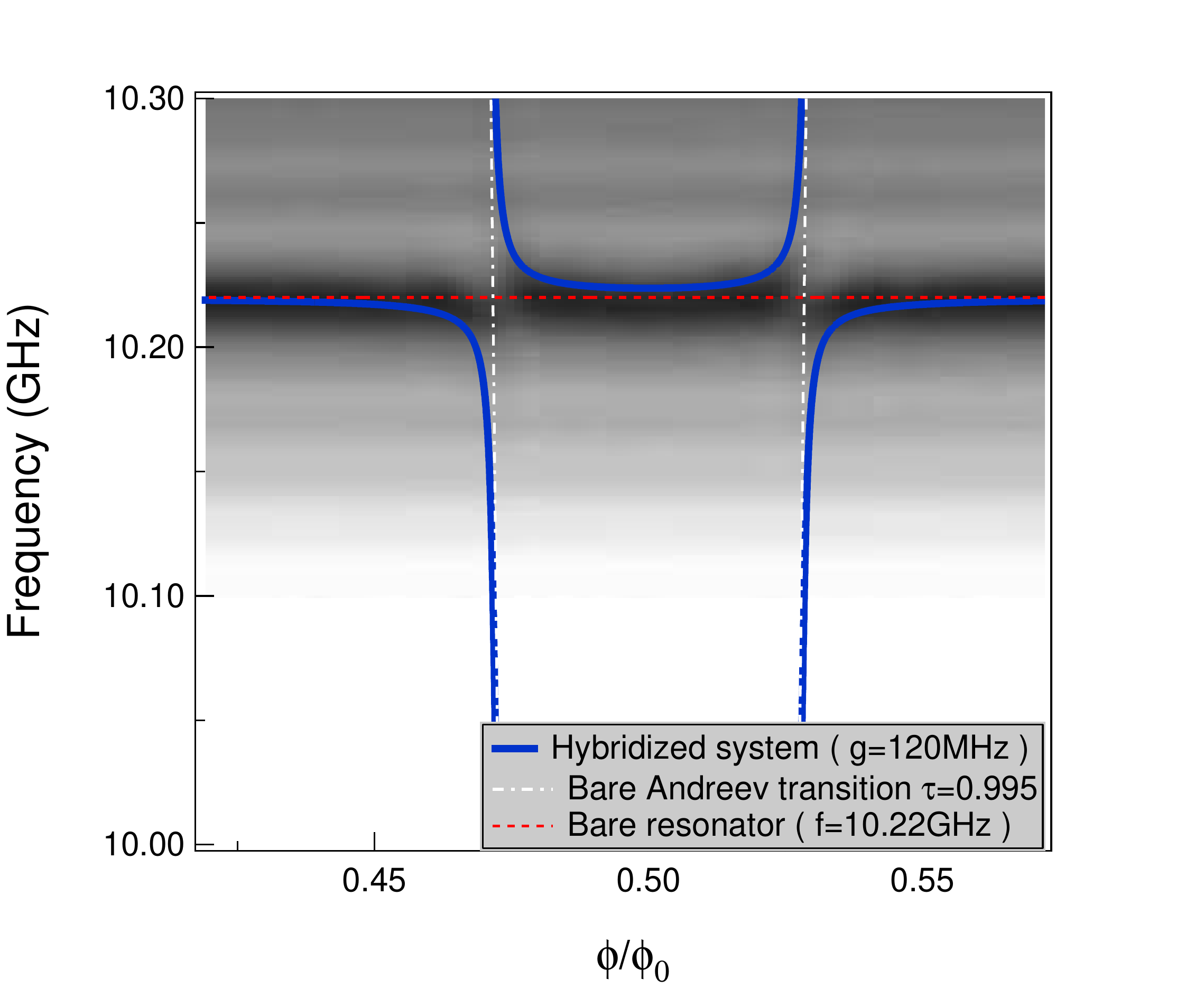}
\par\end{centering}

\caption{One-tone spectrum as a function of probe frequency (vertical axis)
and phase across contact (horizontal). The gray level encodes the
amplitude of the reflection coefficient. The dotted lines correspond
to the bare frequencies for the resonator (red) and the Andreev transition
of a single-channel of transmission $\tau\sim0.995$ (white). The
blue line is the prediction for the coupled system (coupling constant
$g\sim120\,\textrm{MHz}$). }

\label{Full one-tone spectrum}
\end{figure}

\par\end{center}

\subsection{Two-tone spectroscopy}

The spectrum can be explored over a much wider frequency range by
using a two-tone technique. In this case the resonator is constantly
probed at a fixed frequency $\nu_{0}$ close to its resonance $\nu_{R}$,
while sweeping the frequency $\nu_{1}$ of a second tone that is applied
through the directional coupler (see Figure~\ref{Microwave setup}).
The amplitude of this second tone is chopped at an audio frequency
$\nu_{a}$. The reflected signal at $\nu_{0}$ is homodyne detected
yielding the two quadratures I and Q which are then measured by lock-in
amplifiers at $\nu_{a}$ (see Figure~\ref{IQmixer}).

\begin{center}
\begin{figure}[h]
\centering{}\includegraphics[width=12cm]{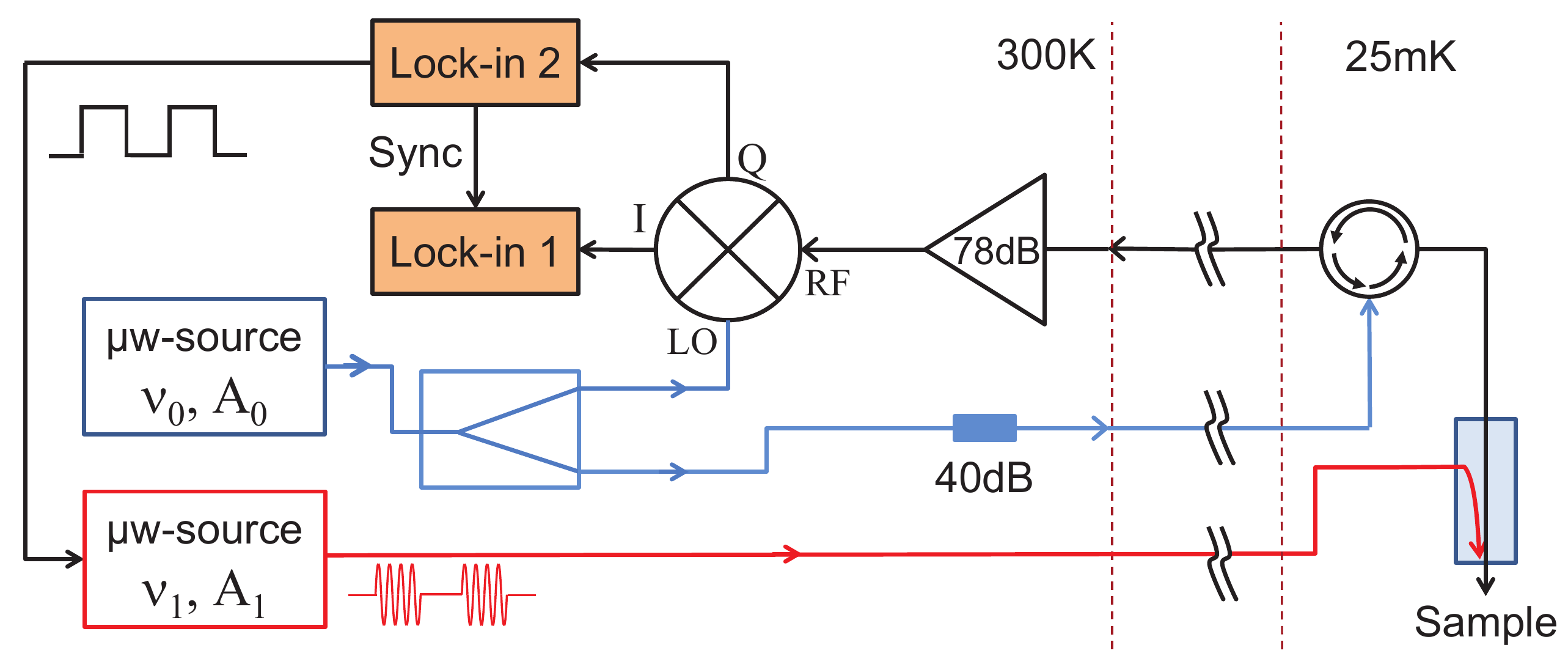}\caption{Room temperature setup for the two-tone spectroscopy: The pump signal
at $\nu_{1}$ is chopped at audio frequency. The reflected signal
at $\nu_{0}$ coming out of the cryostat is amplified and homodyne
detected in a IQ mixer. The two phases are detected by the two lock-ins
at the audio frequency at which the pump signal at $\nu_{1}$ is chopped.}
\label{IQmixer}
\end{figure}

\par\end{center}

The spectra of the reflected signal as a function of the flux through
the loop (horizontal axis) and the frequency $\nu_{1}$ of the second
tone (vertical axis) is shown in Figure~\ref{two-tone} for two different
contacts. Also shown in the figure are vertical cuts of each of the
spectra at half flux quantum $\left(\delta=\pi\right)$, which are
fitted using two lorentzian peaks, all with linewidths below 60\,MHz.
By fitting the spectra using the analytical expression for the Andreev
transition frequency $2E_{A}/h$, one can extract the gap $\Delta$
of the aluminum film and the transmission of the channels. Despite
the apparently similar shapes of the multiple lines, the spectra do
not correspond to contacts with several channels of slightly different
transmissions, as shown by the continuous lines in \ref{two-tone}(b).
Instead, the appearance of several peaks is attributed to the high
microwave power injected in order to acquire sufficient signal.

\begin{center}
\begin{figure}[th]
\centering{}\includegraphics[width=12cm]{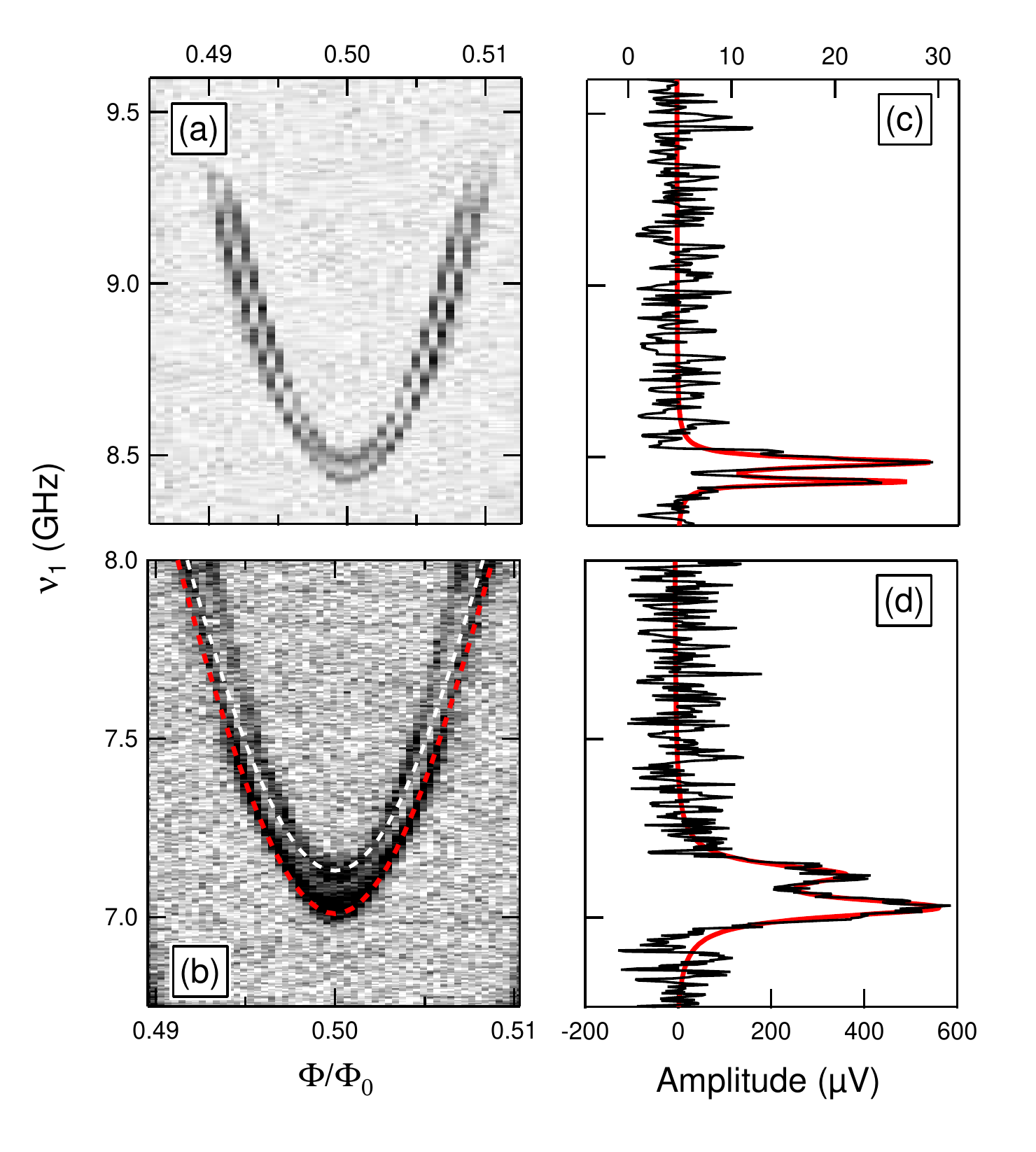}\caption{\textbf{(a)} and \textbf{(b)}: Greyscale coded amplitude of the reflected
signal at $\nu_{0}$, as a function of flux through the SQUID loop
(horizontal) and frequency $\nu_{1}$ of second tone (vertical) for
two different contacts. Fits like red dashed line in (b) using the
analytical expression for the Andreev transition frequency $2E_{A}/h$
determine the channel transmissions ($\tau=0.9\textrm{906}$\textbf{
}in (a) and $\tau=0.9\textrm{934}$ in (b)). The white dashed line
in (b) illustrates the fact that the additional features of the spectrum
cannot be attributed to Andreev transitions of other channels with
slightly different transmission coefficients. \textbf{(c)} Measured
amplitude (black) of reflected signal at a flux $\phi=0.5\phi_{0}$
for contact of panel (a). Red line: fit of measured data using two
lorentzians of widths 17~MHz and 28~MHz. \textbf{(d)} Measured amplitude
(black) of reflected signal at a flux $\phi=0.5\phi_{0}$ for contact
of panel (b). Red line: fit of measured data using two lorentzians
of widths 61~MHz and 65~MHz. }
 \label{two-tone}
\end{figure}

\par\end{center}

\section{Conclusions}

We have presented the first evidence for the efficient inductive coupling
of a superconducting atomic contact to the electromagnetic field of
a coplanar resonator. Using a two-tone setup, we have performed spectroscopy
of the Andreev levels in the contact over several gigahertz. The observed
linewidths are one order of magnitude smaller than in previous experiments
and provide a lower bound of $10\,\mathrm{ns}$ for the coherence
of the Andreev states. Although this is still too short for coherent
manipulation of the Andreev doublet, we have identified a parasitic
heavily damped resonance that loads the coplanar mode in the present
design. A redesign of the resonator geometry to avoid this spurious
resonance is expected to improve the lifetime by an order of magnitude.
Pulsed pump and probe experiments should then allow performing Rabi
oscillations of the state of the Andreev system.

\ack{}{}

We gratefully acknowledge technical support by P. Senat and P. F.
Orfila, and the expert input and assistance of P. Bertet, D. Estève,
P. Joyez and D. Vion. We have also greatly benefited from discussions
with L. Bretheau, A. Levy Yeyati and J. M. Martinis. Financial support
by ANR contracts DOCFLUC and MASH, C\textquoteright{}Nano, the People
Programme (Marie Curie Actions) of the European Union\textquoteright{}s
Seventh Framework Programme (FP7/2007-2013) under REA grant agreement
n\textdegree{} PIIF-GA-2011-298415, and ECOS-SUD (France-Argentina)
n\textdegree{}A11E04 is acknowledged.

\end{document}